\documentclass[iop]{emulateapj}
\usepackage{amsmath}
\usepackage{subfigure}
\usepackage{rotating}
\newif\ifAMStwofonts
\AMStwofontstrue
\def\ga{\mathrel{\hbox{\rlap{\hbox{\lower4pt\hbox{$\sim$}}}\hbox{$>$}}}}
\def\la{\mathrel{\hbox{\rlap{\hbox{\lower4pt\hbox{$\sim$}}}\hbox{$<$}}}}
\shorttitle{Geostationary Orbital Message}

\shortauthors{J. M. Weisberg  \& T. Paglen}

\begin{document}

\submitted{Accepted by Astronomical Journal 2012 August 21}

\title{A Temporal Map in Geostationary Orbit:   The Cover Etching on the EchoStar XVI  Artifact}

\author{Joel M. Weisberg}
\affil{Department of Physics and Astronomy, Carleton College, Northfield, MN 55057; email: jweisber@carleton.edu }

\author{Trevor Paglen}
\affil{Independent Scholar and Artist; email: trevor@paglen.com}
 
\begin{abstract}
Geostationary satellites are unique among orbital spacecraft in that they experience no appreciable 
atmospheric drag. After concluding their respective missions, geostationary spacecraft remain in orbit 
virtually in perpetuity. As such, they represent some of human civilization's longest lasting artifacts. 

 With this in mind, the EchoStar XVI satellite, to be launched in fall 2012, will play host to a time 
capsule intended as a message for the deep future. Inspired in part by the Pioneer Plaque and 
Voyager Golden Records, the EchoStar XVI Artifact is a pair of gold-plated aluminum jackets 
housing a small silicon disc containing one hundred photographs. The Cover Etching, the subject 
of this paper, is etched onto one of the two jackets. It is a temporal map consisting of a star chart, 
pulsar timings, and other information describing the epoch from which EchoStar XVI came. 
The pulsar sample consists of 13 rapidly rotating objects, 5 of which are
especially stable, having spin periods $< 10$ ms and extremely small spindown rates.

In this paper, we discuss our approach to the time map etched onto the cover and the 
scientific data shown on it; and we speculate on the uses that future scientists may have 
for its data.  The other portions of the EchoStar XVI Artifact will be discussed elsewhere. 
\end{abstract}

%\slugcomment{Accepted by Astronomical Journal 2012 August 21}


\section{Introduction}

In the early 1970s, a group of scientists and artists led by Carl Sagan 
and Frank Drake designed several artifacts and messages intended 
for extraterrestrial audiences. The first was the Pioneer Plaque placed 
on the {\it{Pioneer 10}} and {\it{11}} spacecraft, launched in 1972 and 1973, 
respectively (Sagan et al 1972).  In 1974, Frank Drake composed the ``Arecibo 
Message,'' a 1679-bit digital transmission beamed toward star cluster M13 to 
celebrate the renovation of the Arecibo Observatory. More detailed messages, 
including images and sounds,  were encoded onto the phonographic �Golden 
Records� and placed on the {\it{Voyager 1}} and {\it{2}} spacecraft, launched in 1977 toward 
deep space (Sagan et al 1978). 

While geostationary spacecraft do not leave the solar system or even the Earth's 
environs, they inhabit time in a way that is similar to deep space probes. Located 
in the so-called ``Clarke Belt,'' some 36,000 km above the equator, geostationary 
communications satellites experience no appreciable atmospheric drag. These 
spacecraft have one of two fates. International regulations stipulate that satellite 
operators conduct end-of-life maneuvers to place derelict satellites into 
supersynchronous ``graveyard'' orbits (Collis 2009). In practice, however, operators 
regularly leave their spent satellites in loosely geosynchronous orbits. In both 
cases, these spacecraft remain orbiting virtually in  perpetuity (Walker \& King-Hele 
1986; Flohrer et al 2011).  As such, 
geostationary satellites are undoubtedly some of contemporary civilization's 
longest-lasting objects. As a scientific and cultural exercise, we can imagine 
scientists in the distant future looking in this zone for relics from past 
civilizations (Freitas \& Valdes 1983, 1985). The EchoStar XVI Artifact represents 
an effort to communicate with these future scientists.   

The full Artifact is composed of two interlocking gold-plated aluminum jackets 
housing a silicon disc called ``The Last Pictures,'' on which 100 photographs are 
nano-etched. The Artifact Cover 
Etching is located on the exterior of one of the two jackets. 

In this paper, we focus on the Artifact Cover Etching, wherein we 
endeavored to create a map explaining {\it{when}} the spacecraft originated, 
to whomever may find it in the future.  Specifically, we detail the several 
scientific and mathematical messages we developed for the Artifact Cover.
The  photographic  contents of the Artifact's silicon disc are detailed elsewhere 
(Paglen 2012). 

\section{Artifact Cover  Theory and Design}

Some of the basic ideas that informed our time-map were developed by 
Sagan et al (1972, 1978) for the {\it{Pioneer}} and {\it{Voyager}} missions, which we 
took as a starting point. Scientific developments since the 1970s have provided 
some interesting new possibilities described below.  Another essential difference 
between this and earlier plaques is that ours is locked in perpetual Earth orbit, and 
therefore obviously associated with our planet.  Consequently, we made no effort to 
communicate the Artifact's {\it{place}} of origin. Instead, our temporal map seeks to indicate 
the spacecraft's {\it{Epoch}} of Origin using {\it{currently}} measured quantities such as the  
South Celestial Pole location, Earth's stellar rotation period, location of tectonic plates, 
orbital period of the Moon, locations of stars and extragalactic radio sources, and positions 
and periods of pulsars. The map, as etched onto the Artifact Cover, is shown in Fig. 1, 
and its various elements are discussed below.

\subsection{Fundamental Unit of  Frequency, Time, and Epoch of Origin}

We use the hyperfine ground-state transition of neutral atomic hydrogen as the fundamental 
unit of frequency  $f_0$ and time $t_0 = 1 / f_0$ on the Artifact Cover. This transition is illustrated 
by a sketch  of neutral atomic hydrogen in each 
of the two possible hyperfine states  (see Fig. 1).  The sketch is identical to the one on the 
Pioneer Plaques, which used the same transition as their
fundamental temporal unit (Sagan et al 1972). Although the International Committee for 
Weights and Measurements pegs 
the duration of one second to the ground-state cesium-133 hyperfine transition 
(Conference Generale des Poids et Mesures 1968), we chose to continue using the hydrogen 
transition both as a nod to the work of Sagan et al (1972), in addition to the following practical 
reasons: Hydrogen is the most abundant element in the universe; and, as the simplest element,
 it is far easier to sketch than cesium. Equally important, the currently accepted value of the 
 hydrogen hyperfine transition, with $f_0  = (1,420,405,751.768 \pm 0.001)$ s$^{-1}$  
 (Kramida 2010), is adequately defined  and sufficiently accurate and precise for our purposes. 
 
 The Epoch of Origin is nominally 2011 Oct 4.0 UT (MJD 55838.0). However, some of the items 
 on the Artifact Cover were plotted using a slightly different epoch for ease of calculation, if the results
 are not noticeably affected.  The actual epochs are given below.

\subsection{Fundamental Spatial Reference System}

The Artifact Cover requires a  spatial reference system against which to measure
various quantities.  A plot of the locations of the brighter stars and pulsars (see \S \ref{sec:msp})
centered on the South Celestial
Pole (SCP), with the current location of the SCP also marked, serves adequately on a timescale of 
 $\lesssim 10^3$ yr.  On longer timescales, the SCP and many of the stars and pulsars will move in 
 various directions  across the sky, rendering their patterns on the Artifact Cover increasingly hard 
 to decipher.   (See \S \ref{sec:prec} for discussion of the future motions of the SCP.)

Consequently, we require longer-lived, relatively fixed objects  or phenomena to delineate
 the reference frame on the longest timescales.  Therefore we also plotted the brightest point-like 
 radio sources  (NRAO  2011).  These objects, mostly Active Galactic Nuclei (AGN) in deep space, 
 will exhibit very small angular motions over cosmic timescales, although their intense radio emissions 
 will  fade within $10^{7-8}$ yr (Porciani et al 2004).

\subsection{Epoch of Origin Markers}

In this subsection, we detail each of the Epoch of Origin markers located on the Artifact Cover.  
As we shall later show, the redundant nature of the information on these Markers enables them
to be used for other purposes as well.

\subsubsection{Terrestrial Spin Axis Precession}

\label{sec:prec}
The precession of the Earth's spin axis provides one means of identifying the Epoch of Origin 
of this Artifact.  The intersection of the two lines crossing the sky map in Fig. 1 marks the current 
location of the SCP\footnote{The lines themselves mark the four cardinal 
points of {\it{RA}}:  0, 90, 
180, and 270 degrees; for Epoch of  Origin = 2000.0 CE.}. Note however that the approximately 
cyclical nature of the  
precession  (with $23\fdg3 \pm 1\fdg3$  obliquity)  brings the SCP close to this location at 
$\sim26$-kyr 
intervals,  rendering precession unusable as an Epoch of Origin Marker on timescales of this order 
or longer. On very long timescales, however, Earth's spin axis obliquity may undergo major  
chaotic changes due to solar system resonances (Tomasella et al 1996,  Neron de Surgy \& Laskar 
1997), at which time  the marked location and hence the obliquity of 
today's pole would again serve as an Epoch of Origin marker, albeit crude.

\subsubsection{The Earth's Stellar Day and the Moon's Sidereal Month}

According to  Lunar Laser Ranging, our Moon's orbital radius is expanding by 38 mm/yr 
(Chapront et al 2002), at the expense of a lengthening of the Earth's spin period.  Hence, by   
specifying on Fig. 1 the Earth's stellar\footnote{The Earth's  ``stellar rotation period'' or ``stellar
 day'' is measured with respect to the stellar inertial reference frame.  This quantity is slightly 
different than the sidereal day, which is measured with respect to the (precessing) equinox.}  
rotation period to be\footnote{One Earth stellar rotation period $\equiv 1$ Earth stellar day 
=  $2 \pi /  \Omega$ with $\Omega= 7.292 115 0(1) \times 10^{-5}$ rad/s measured over the 
period 1978-1994 CE (Groten 1999).}  86,164.101 s, and the Moon's sidereal orbital 
period [1 sidereal month = 27.321661 mean solar days = 27.396462 stellar days (Kay \& Laby 1995) ], 
we  implicitly specify the epoch at which these constantly-changing quantities were measured 
(approximately 1990 CE)\footnote{The value of the sidereal month, found next to the crescent Moon 
on the etching, is written in units of stellar days rather than in units of $t_0$, in order to avoid the need 
for 29 trailing zeros.  This is the only temporal quantity on the etching {\it{not}} specified in units of $t_0$.}

\subsubsection{Tectonic Plate Motions}

The map of the Earth itself provides a much cruder estimate of the Epoch of Origin of the Artifact, 
since motion of the tectonic plates will slowly distort the shape and location of the continents over time.

\subsubsection{Millisecond Pulsars}

\label{sec:msp}
By far the most precise Epoch of Origin markers are provided by the inscribed pulse periods of 
 millisecond pulsars (MSPs).   %MSP are rapidly spinning (up to ~700 rotations / s) neutron stars 
 %whose beamed radio emission points at us for an instant during each rotation. 
Since all pulsars' intrinsic  spin periods are lengthening with time, the choice of a particular value 
of the period specifies the epoch at which it was determined.  Pulsars, having been discovered  
 only a few years before the Pioneer launches (Hewish et al 1968), were also used as Epoch 
 (and Location) of Origin Markers on those spacecraft plaques.  However, although the pulsars 
 known at that time were remarkably stable clocks, they can not compare with the stunning stability 
 (Shannon \& Cordes 2010, Hartnett \& Luiten 2011)  of the subsequently discovered (Backer et al 1982)
  MSPs.  We determined $P(T)$, the MSP period at our Epoch of Origin $T$, using its measured 
   period $P_0$ at a given epoch  $T_0$, and period derivative or 
  ``spindown rate''  $\dot{P}$ (Manchester et al 2005):
\begin{equation}
P(T)=P_0 + \dot{P} (T - T_0).
\end{equation}
The period uncertainty $\Delta P$ and Epoch of Origin uncertainty $\Delta T$ 
were calculated by propagating the uncertainties (Manchester et al 2005) in the
 quantities used in Eq. 1:
\begin{equation}
\Delta T = \frac{\Delta P(T)}{ \dot{P}  } = \frac {  \sqrt {   [\Delta P_0]^2 +
[\Delta \dot{P} (T - T_0) ]^2   }      }{ \dot{P}  }.
\end{equation}

While normal pulsars are thought to stop shining after $10^{6-7}$ yr,  MSPs, which have been 
spun up by accretion of matter from a companion, appear to have much longer radio-emitting 
lifetimes (possibly comparable to the age of the universe).   Table 1 and Fig. 1 give the locations 
and periods  of 13 MSPs at Epoch of Origin $T =$ 2011 Oct 4.0. (Figure 1  gives the pulsar 
periods in 
units of $t_0$, using base-2 numbers.)  
The Table also lists $\Delta T$,   the remarkably small uncertainty in $T$, calculated via Eq. 2.  
The final set of 13 MSPs was winnowed from the list of all pulsars in the ATNF 
catalogue (Manchester et al 2005) 
as follows:  First, we eliminated pulsars with periods $P > 0.100$ s. All pulsars in globular 
clusters 
were also eliminated because their circumcluster orbits lead to complicated, 
temporally-varying changes 
in their apparent periods.  Next, we removed those with measured angular 
velocities across the sky, since 
such motions, if measurable,  are also large enough to render the objects invisible 
or unidentifiable on long 
timescales.  Then,  short orbital-period binary pulsars were eliminated because radio 
beams may precess away from 
Earth  due to the pulsars' relativistic interactions with the companion (Weisberg \& 
Taylor 2002), or to pulsars' 
demise because of gravitational radiation-induced orbital decay 
(Weisberg et al 2010) and coalescence.  
Some of the final 13 MSPs as selected above are actually young pulsars with rather large 
spindown rates and short 
lifetimes.  The periods that are written on the Cover Etching are truncated just beyond 
the last significant (base-2) 
digit as indicated by Eq. 2, in accord with standard scientific practice.  Additional MSPs, 
though meeting the above 
criteria, are plotted without written periods simply because the string of 
binary digits would not fit comfortably on the Artifact Cover. 

\section{Future Fate of the Artifact and its Cover}

At the end of its active life approximately 15 years after launch, EchoStar XVI is 
scheduled to boost into a 
slightly higher ``graveyard'' orbit in order to free up valuable space in the geostationary 
zone.   Interestingly, 
as the Earth's spin gradually slows, the altitude of the geostationary zone will migrate 
outwards, eventually 
to coincide with today's graveyard.  The spacecraft and its Artifact should orbit 
indefinitely in the graveyard, 
since its altitude places it well above the last wisps of atmosphere.   They will 
occasionally encounter space 
debris (both natural and human-made) in this zone, but the Artifact's hardened 
nature should shield it from 
all but the most extreme collisions.  According to recent modeling (Schr{\"o}der  \&
Connon Smith 2008), the Sun's luminosity will slowly increase over the next 5 billion yr 
to almost double its 
current value, thereby also doubling the solar flux delivered to the Artifact, which is 
designed easily to 
survive the enhanced flux.  At around the same time, fusion in the Sun's core will 
cease due to hydrogen depletion, 
and over the next 2 billion years its outer layers will expand approximately to the 
radius of the Earth's orbit, thereby 
enveloping (or nearly so) our planet and its surroundings. (Earth's orbit will also 
migrate outwards as the Sun loses 
significant mass due to winds at these late stages, which may prevent our direct 
immersion.)  At that time, the Sun 
will also radiate almost three thousand times its current luminosity.  The Artifact, 
even if it has survived to this point, 
will likely not outlive these events.

\section{Discussion}

A comparison of the messages and the technologies on the current Artifact with those 
on the Pioneer and Voyager 
spacecraft indicates the rapid pace of technological advance over even just a few 
decades.  While we have 
endeavored to use putatively universal symbols, the question of whether our temporal 
map would be decipherable 
by deep-future humans or extra-terrestrial spacefaring civilizations is a deeply controversial 
one that intersects fields 
from philosophy of science to anthropology, semiotics, and cognitive science. As 
such, the question of its future intelligibility  
is outside the scope of our discussion here. To illustrate how future scientists and 
archaeologists may use the data provided 
in our time-map, however, we will assume that it will be intelligible by any spacefaring 
civilization; i.e., any civilization with 
the means to examine it.

\subsection{Determining the Time Elapsed between Origin and Discovery}

Our Etching's temporal map contains a mixture of ``clocks,'' some of which are 
expected to run at constant rates over 
cosmological times (e.g., the hydrogen hyperfine transition frequency), while others' 
rates will vary (e.g., the Earth's 
spin period).  Depending  on the level of their scientific knowledge, the discoverers 
may be able to determine the Epoch 
of Origin of the Artifact by comparing Epoch of Origin  clock rates with Epoch of 
Discovery  clock rates.  For example, 
they could measure at the Epoch of Discovery, the periods $P$ and spindown 
rates   $\dot{P}$
of some of the 13 MSPs whose Epoch of Origin periods are written on the etching.  
Then they could use Eq. 1 to determine $(T-T_0),$
the time elapsed between the two epochs.  

If the elapsed time is sufficiently long, 
higher derivatives must be added to Eq. 1.  While current pulsar 
theory gives an expected value of the next derivative $\ddot{P}  $ under the 
assumption of rotational energy loss to magnetic dipole radiation (Lyne \& Graham-Smith 
2012, p. 72),
\begin{equation}
\ddot{P}= - \frac{\dot{P}^2}  {P},
\end{equation}
this expression has been shown to be a poor  
approximation to reality in the few cases where the next derivative has actually been 
measured (Gradari et al 2011).  
Similar arguments apply to the other �clocks� such as the Earth's spin period and the 
Moon's orbital period.  Nevertheless, the discoverers should be  able to ascertain the true 
elapsed time by checking the consistency of their 
multiple results from an  ensemble of such ``clocks.'' 

\subsection{Testing and Refining Scientific Theories} 

Once the discoverers  establish $(T-T_0),$ the time elapsed since the origin of the Artifact, 
they  will also have the opportunity  to test and refine  some of their 
scientific theories over deep time.   For  example, they could combine our 
inscribed data with their measurements, thereby directly determining 
higher order derivatives as discussed above.  The assumptions underlying the theories
could then be tested via comparison
of these measured higher derivatives with the theoretically predicted ones,  
as shown in the above pulsar spindown example.  

Some of our current clocks may ``break'' over cosmic timescales, in which case 
the discoverers will derive
 totally unique information by decoding our measurements.  For example, the pacing of 
 resonances causing major 
 changes in our spin axis obliquity is dictated by changes in tidal dissipation  in the 
 Earth-Moon system, which is 
 dependent on tectonic plate motions and other geophysical processes 
 whose details are poorly known over long
  timescales, but which could be constrained by data from our era.  
  
  The discoverers  might further be 
  able  to apply our currently recorded  measurements for uses that we 
  cannot foresee, much as we now use ancient solar eclipse records to provide unique 
  data on the evolution of the 
  Earth's spin period over historic times (Stephenson 1997).  In addition, the mere 
  discovery of the Artifact and  
  spacecraft in geostationary space could provide the discoverers with a wealth 
  of information on us, much as the 
  discovery of the ``Antikythera Mechanism'' astronomical computer (Freeth et al 2006) 
  in a shipwreck 
  has done for our knowledge of 
  ancient Mediterranean scientific and technological expertise. Further, we can hope 
  that any collisions suffered by 
  the Artifact before its discovery will not prevent its decoding, much as erosion of the 
  Antikythera Mechanism slowed 
  but did not stop current researchers from deciphering much of it.

\acknowledgements{The idea for this work was created by TP.  JMW acknowledges  financial
support from a Eugster 
Fellowship; and  thanks J. E. Everett for useful comments, and the late Philip Morrison, SETI pioneer 
(Cocconi \& Morrison 1959) and mentor, for inspiration.}

\newpage
%%%%%%%%%%%%%%%%%%%%%%%%%%%%%%%
\begin{deluxetable}{llrl}
\tablecolumns{4}
\tablecaption{Table 1:  Parameters of the millisecond pulsars whose periods are 
inscribed on the Cover Etching.}
\tablehead{
\colhead{Pulsar Name}  &  \colhead{Period $P$ at $T =$} & \colhead{$\dot{P}$}  & \colhead{$\Delta T$}    \\
\colhead{}                          & \colhead{2011 Oct 4.0 UT (s)}    &  \colhead{s/s} &  \colhead{(d)}    \\
}
\startdata
J1016-5819  &	0.0878344385(2)           & $6.980(4) \times 10^{-16}$  & 2.7      \\
J1629-6902  & 0.0060006034468(1)    & $1.00(3) \times  10^{-20}$  & 127      \\
J1454-5846  & 0 .045248773318(4)      & $8.17(9) \times  10^{-19}$  & 50      \\
J1435-6100  & 0.0093479722199(2)    & $2.45(5) \times  10^{-20}$  & 93      \\
J1420-6048  & 0.068210329(1)             & $8.3167(3) \times 10^{-14}$  & 0.15      \\
J1232-6501  & 0.08828190855(1)        & $8.1(3) \times 10^{-19}$  & 169      \\
J1216-6410  & 0.00353937565876(4) & $1.6(2) \times 10^{-21}$  & 288      \\
J1125-6014  & 0.00263038074072(2) & $4.01(9) \times 10^{-21}$  & 61      \\
J1112-6103  & 0.0649748526(5)          & $3.1460(1) \times 10^{-14}$  & 0.2      \\
J1055-6028  & 0.09966459910(6         & $2.95322(5) \times 10^{-14}$  & 0.03      \\
J1103-5403  & 0.00339270965950(6) & $3.7(10) \times 10^{-21}$  & 192      \\
J0940-5428  & 0.0875586849(4)          & $3.2868(1) \times 10^{-14}$  & 0.14      \\
J0855-4644  & 0.06468893573(1)        & $7.26269(3) \times 10^{-15}$  & 0.02      \\
\enddata 
\tablecomments{Uncertainties in the last digit(s) of measured quantities are given in parentheses..}
\end{deluxetable}
%%%%%%%%%%%%%%%%%%%%%%%%%%%%%%%

\newpage

%%%%%%%%%%%%%%%%%%%%%%%%%%%%%%%%%%%%%%%%%%%%%%%%

\begin{figure}

\caption{
%\footnotesize
\small
The Cover Etching portion of the Artifact, to be launched on the exterior of the 
EchoStar XVI satellite. The image is dominated by a map of the southern sky.    
Stars are denoted by circles whose size is proportional to their apparent brightness 
at visual wavelengths.  Point  radio sources, mostly Active Galactic Nuclei, are represented 
by squares.  Millisecond pulsars are symbolized by triangles; some of whose periods on Epoch of
Origin Date 2011 Oct 4.0 UT are given in base-2 numbers, which utilize only zeros (``-'') 
and ones (``$|$''),and a placeholder (``$\times$'')  to distinguish the whole and fractional parts.  The 
two lines crossing the whole skymap denote RA = (0, 90, 180, and 270)$\degr$, and their 
crossing point marks the South Celestial Pole at Epoch  2000.0 CE. Also shown are the current 
arrangement of continents on the Earth, the rotation period of the Earth and the revolution period of 
the Moon.  All time intervals are given in units of the period of the ground-state hyperfine transition 
of atomic hydrogen, except for the lunar sidereal orbital period, which is  denoted in units of the 
stellar day so as to avoid the need for 29 trailing zeros. Practice with our binary number system is 
given by presenting two values next to geometrical objects:   the ratio of the hypotenuse and the 
side of a (displayed) $45\degr$ right triangle;  and the ratio of the areas of  a circle to its 
circumscribed square (also displayed). }

\includegraphics[trim=2.5in 0in  0 0.5in ,clip,scale=0.6]{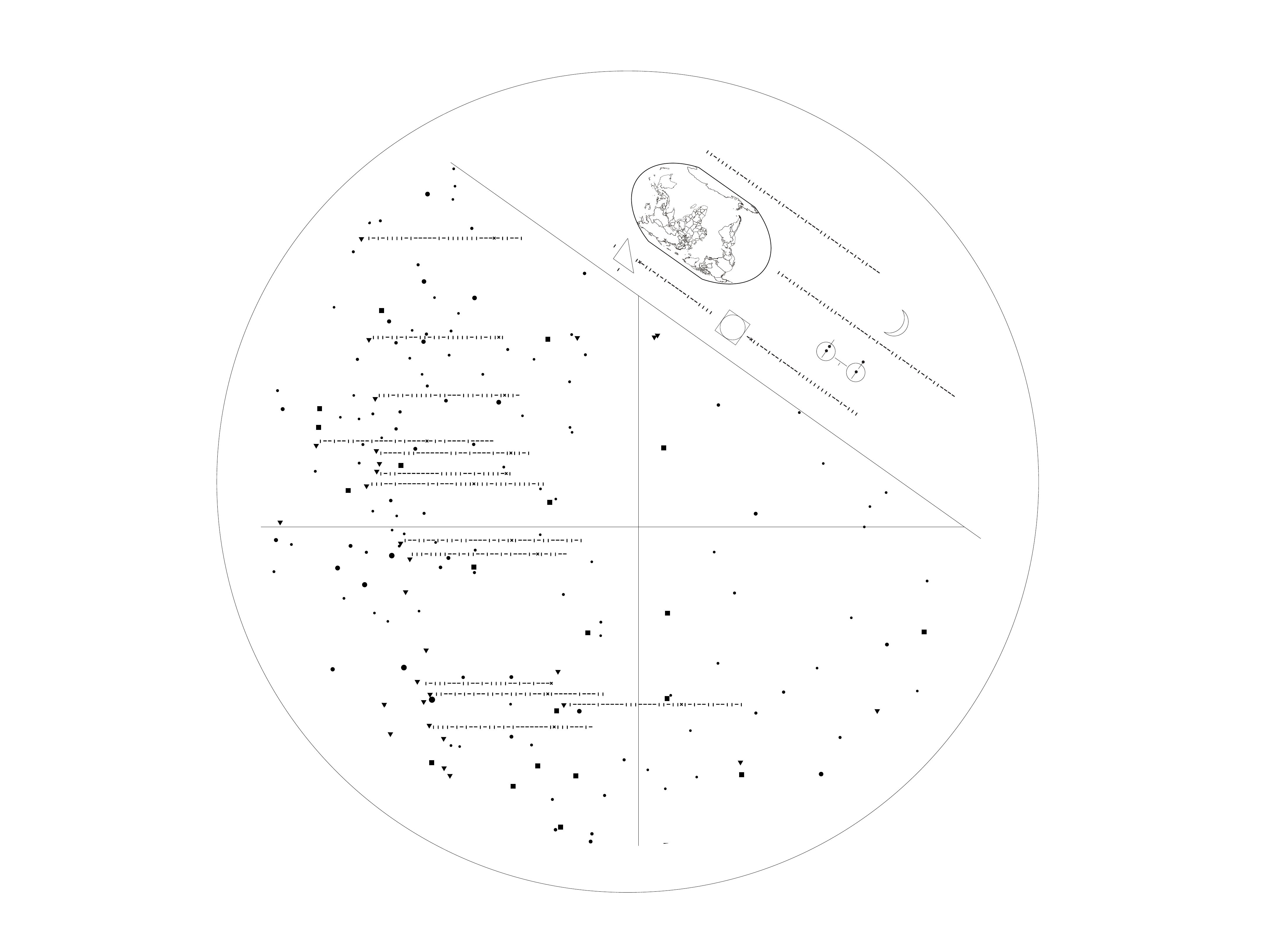}
\end{figure}
%%%%%%%%%%%%%%%%%%%%%%%%%%%%%%%%%%%%%%%%%%%%%%%%

\end{document}